\documentclass[aps,showpacs,showkeys,prb,preprint,amsmath,amssymb]{revtex4}
\usepackage{amssymb}
\usepackage{amsmath}
\usepackage[dvips]{graphicx}

\begin{document}
\title {$Fe_{0.5}Cu_{0.5}(Ba_{1-x}Sr_{x})_2YCu_2O_{7+\delta}$ ($x$ = 0, 0.5 and
1) superconductors prepared by high pressure synthesis}

\author{Z.A. Ren}
\email[Corresponding author's email: ]{renzhian@ssc.iphy.ac.cn}
\author{G.C. Che}
\author{Y.M. Ni}
\author{C. Dong}
\author{H. Chen}
\author{S.L. Jia}
\author{Z.X. Zhao}
\affiliation{National Laboratory for Superconductivity, Institute of Physics,
Chinese Academy of Sciences, P.O.Box 603, Beijing, 100080, P.R. China}
\date{\today}

\begin{abstract}
In this paper, the $Fe$-containing superconductors
$Fe_{0.5}Cu_{0.5}Ba_2YCu_2O_{7+\delta}$,
$Fe_{0.5}Cu_{0.5}BaSrYCu_2O_{7+\delta}$ and
$Fe_{0.5}Cu_{0.5}Sr_2YCu_2O_{7+\delta}$ were successfully prepared by common
solid-state reaction followed with a procedure of high pressure synthesis. The
structural change and superconducting properties in
$(Fe_xCu_{1-x})BaSrYCu_2O_{7+\delta}$ ($x$ = 0 $\sim$ 1.0) systems were also
investigated. Annealing experiments indicate that the occurrence of
superconductivity in $Fe_{0.5}Cu_{0.5}(Ba_{1-x}Sr_{x})_2YCu_2O_{7+\delta}$ ($x$
= 0, 0.5 and 1) systems is mainly induced by the procedure of high pressure
synthesis, which causes the increase of oxygen content and the redistribution
of $Fe$ atoms between $Cu(1)$ and $Cu(2)$ sites, but not from possible
secondary phase of $YBa_2Cu_3O_{7-\delta}$, $YBaSrCu_3O_{7-\delta}$ or
$YSr_2Cu_3O_{7-\delta}$ superconductors.
\end{abstract}


\maketitle


\section{Introduction}

It is well known that $YBa_2(Cu_{1-x}M_x)_3O_{6+\delta}$ ($M$ = $Fe$ and $Co$)
systems (We denote it as $(Fe_xCu_{1-x})Ba_2YCu_2O_{7+\delta}$) have been
extensively investigated for providing information about the mechanism of
superconductivity and understanding the correlation among superconductivity,
magnetism and crystal structure in high temperature ceramic superconductors
\cite{30,31,32,33,34,35,36,37,38,39,40,41,42,43,44,45}. The main conclusions
concerning its crystal structure and superconducting properties are as follows:

\setcounter{paragraph}{0}
\paragraph{}
In undoped $YBa_2Cu_3O_{7-\delta}$ superconductor, there are two different
structural $Cu$ sites: $Cu(1)$ and $Cu(2)$. The $Cu(1)$ site has a square
planar oxygen coordination and forms $Cu-O$ chain, and the $Cu(2)$ site has a
five fold pyramidal coordination of oxygen and forms $CuO_2$ plane. There is
strong evidence from neutron diffraction results that $Cu(1)$ sites are the
preferred occupation sites for $Fe$ and $Co$ atoms in substituted
$(Fe_xCu_{1-x})Ba_2YCu_2O_{7+\delta}$ compound \cite{30,31,32,33,34}.

\paragraph{}
Neutron and X-Ray diffraction analysis indicate that
$(Fe_xCu_{1-x})Ba_2YCu_2O_{7+\delta}$ undergoes a structural phase transition
from orthorhombic to tetragonal at $Fe$ concentration $x$ $\sim$ 0.12$-$0.15
\cite{36,37,38}. Electron microscopy observations indicate that the
microstructure of $(Fe_xCu_{1-x})Ba_2YCu_2O_{7+\delta}$ evolves as the $Fe$
concentration changes \cite{39,40,41,42,43}.

\paragraph{}
The superconducting transition temperature $T_c$, decreases with the increasing
$Fe$ concentration $x$, when $x$ exceeds 0.3, superconductivity disappears
completely \cite{30, 35, 43, 45}.

\paragraph{}
The oxygen content, $7+\delta$, in $(Fe_xCu_{1-x})Ba_2YCu_2O_{7+\delta}$,
increases with the increasing $Fe$ concentration due to the higher valence of
$Fe$ \cite{30, 43, 44}.

In previous studies, several groups \cite{12,13,14,15,16,17,18} have reported
superconductivity in $YSr_2Cu_{3-x}M_xO_{6+\delta}$ compounds with $Fe$ (or
$Co$) light-substitution, but the heavily-substituted compounds ($x$ $>$ 0.3)
did not exhibit superconductivity. T. Denetial \cite{19} observed
superconductivity with $T_c$ $\sim$ 30$-$50K in $Fe$($Co$ or $Ti$)-doped R-123
phase $YSr_2Fe_xCu_{1-x}O_{6+\delta}$ with $x$ = 0.3, while $x$ $>$ 0.3,
superconductivity could not be observed. F. Shi {\it et al.} \cite{20} improved
the superconductivity of $YSr_2Cu_{2.7}Fe_{0.3}O_{7+\delta}$ compound by
high-pressure oxygen annealing, $T_c$ reaches 60K and a shielding fraction of
nearly 100$\%$ is achieved. Recently J. Shimoyama {\it et al.} \cite{21}
prepared $FeSr_2YCu_2O_{7+\delta}$ superconductor with $T_c$ $\sim$ 60K using a
complex synthesis procedure and T. Mochiku {\it et al.} \cite{22} determined
the crystal structure of this superconductor by neutron powder diffraction
studies. Besides, some $Fe$-containing oxides with perovskite structure, such
as $(Pb_{0.5}Fe_{0.5})Sr_2(Y_{0.5}Ca_{0.5})Cu_2O_7$ \cite{23},
$BaR(Cu_{0.5+x}Fe_{0.5-x})_2O_{5+\delta}$ ($R$ = $Y$, $Sm$) \cite{24, 25, 26},
$Bi_2Sr_3Fe_2O_x$ \cite{27}, $BaYCuFeO_5$ \cite{28},  $etc$. have also been
reported, but none of these compounds is superconducting.

We have prepared a $Fe$-containing cuprate superconductor
$Fe_{0.5}Cu_{0.5}BaSrYCu_2O_{7+\delta}$ with $T_c$ $\sim$ 77K in year $2000$ by
solid-state reaction and high pressure synthesis. The preliminary results of
this superconductor have been reported\cite{29}. After that, the studies of
$Fe$-containing superconductors were developed and new superconductors of
$Fe_{0.5}Cu_{0.5}Ba_2YCu_2O_{7+\delta}$ and
$Fe_{0.5}Cu_{0.5}Sr_2YCu_2O_{7+\delta}$ were successfully prepared. In this
paper, the preparation, superconductivity, structure and the results related to
these superconductors are presented.

\section{Experiment}

\subsection{Sample Preparation}

The predried $Y_2O_3$, $BaCO_3$, $SrCO_3$, $Fe_2O_3$, and $CuO$ powders of
99.99$\%$ purity were used as the starting materials. The powders with the
stoichiometric composition of
$Fe_{0.5}Cu_{0.5}(Ba_{1-x}Sr_x)_2YCu_2O_{7+\delta}$ ($x$ = 0, 0.5, and 1.0)
were mixed, ground thoroughly and calcined twice at 925$^o$C for 60 hours in
air with intermediate grinding. The products were pressed into pellets, and
calcined again at 925$^o$C for 60 hours and cooled down to room temperature at
the rate of 30$^o$C per hour in air. These samples prepared by the common
solid-state reaction procedure were labeled as AM-sample.

The AM-sample powders were oxygenated under high pressure of 6GPa at 1000$^o$C
for 0.5 hour by the addition of $5wt.\%$ $KClO_4$ (which was used as an oxygen
source) in a six-anvil of tungsten carbide high pressure apparatus. Samples
were quenched from high temperature quickly by cutting off the furnace power
before releasing the high pressure\cite{29}. These samples were labeled as
HP-sample.

\subsection{Sample Characterization}

The phase and structure of these samples were characterized by powder X-Ray
diffraction (XRD) analysis on an $MXP18A-HF$ type diffractometer with
$Cu$-$K_{\alpha}$ radiation. All investigated samples exhibited single phase
diagram. $PowderX$, $Finax$ and $Rietweld$ programs were used for lattice
parameter calculations. The data of DC magnetization and electrical resistance
were measured using a DC-SQUID magnetometer (Quantum Design MPMS 5.5T) and
standard four-probe technique respectively.

\section{Experimental Results}

\subsection{The structural change and superconductivity in
$(Fe_xCu_{1-x})BaSrYCu_2O_{7+\delta}$ synthesized by solid-state reaction under
ambient pressure}

Fig.1 shows the XRD patterns of the $(Fe_xCu_{1-x})BaSrYCu_2O_{7+\delta}$
systems of AM-samples ($x$ = 0$\sim$1.0), indicating that all samples are of
single phase. Fig.2 shows the changes of lattice parameters. With the
increasing of $Fe$ concentration $x$, lattice parameter $a$ increases and $b$
decreases, and a structural transition undergoes from orthorhombic to
tetragonal at $Fe$ concentration $x$ $\sim$ 0.15, afterwards lattice parameter
$a$ increases slightly. The lattice parameter $c$ always decreases whereas unit
cell volume $V$ increases with the increasing $x$. Fig.3 shows some typical
curves of resistivity $vs.$ temperature, which indicates that the substitution
of $Fe$ atoms suppresses the electric conductance and superconductivity for the
AM-samples, when $x$ $>$ 0.3, all samples become non-superconducting and
exhibit semi-conducting behavior in $R$-$T$ curves. The inset is the dependence
of superconducting transition temperature on $Fe$ content $x$, which displays a
linear depression. These results are similar to those of
$(Fe_xCu_{1-x})Ba_2YCu_2O_{7+\delta}$ system.

\subsection
{Superconductivity of $(Fe_xCu_{1-x})BaSrYCu_2O_{7+\delta}$ prepared by high
pressure synthesis}

XRD patterns of $(Fe_xCu_{1-x})BaSrYCu_2O_{7+\delta}$ HP-samples indicate that
all samples ($x$ varies from 0 to 1.0) are still single phase after high
pressure synthesis. Fig.4 shows $T_c(onset)$ $vs.$ $Fe$ concentration $x$ and
magnetization $vs.$ temperature curves obtained using ZFC mode under applied
external magnetic field of 10Oe, indicating that all samples exhibit
superconductivity and the samples with $x$ = 0.35$-$0.6 have relatively high
superconducting transition temperature $T_c$ and high superconducting volume
fraction $V_m$. Typical curves of the dependence of resistivity and
magnetization on temperature for $Fe_{0.5}Cu_{0.5}BaSrYCu_2O_{7+\delta}$
superconductor are shown in Fig.5, it can be calculated from the ZFC and FC
curves that the superconducting shielding volume fraction is 48$\%$ and the
Meissner volume fraction is 31$\%$ at 10K by the relationship of $V_m$ =
$(4\pi\rho M/H)$, where $\rho$ is the density of sample in $g/cm^3$, $M$ is
mass magnetization in $emu/g$ using ZFC and FC data respectively, and $H$ is
the applied magnetic field in $Oe$.

\subsection{$Fe_{0.5}Cu_{0.5}Ba_2YCu_2O_{7+\delta}$ superconductor}

As mentioned previously, the $Fe_{0.5}Cu_{0.5}Ba_2YCu_2O_{7+\delta}$ sample
prepared by solid-state reaction in air is semi-conductor. While after high
pressure synthesis, this sample becomes superconducting. Fig.6 presents the XRD
patterns of the AM-sample and HP-sample, $R$-$T$ and $M$-$T$ curves of
HP-sample. The lattice parameters for HP-sample are $a$ = 0.3870(3)nm, $c$ =
1.1601(5)nm, while for AM-sample, $a$ = 0.3871(3)nm, $c$ = 1.1671(5)nm. From
the $R$-$T$ curve, the superconducting transition temperature, $T_c(onset)$ is
found at 83K and $T_c(zero)$ is at 63K. From the ZFC and FC curves it can be
calculated that the superconducting shielding volume fraction is 55$\%$ and the
Meissner volume fraction is 22$\%$ at 10K.

\subsection{$Fe_{0.5}Cu_{0.5}Sr_2YCu_2O_{7+\delta}$ superconductor}

The synthesis and crystal structure of $FeSr_2YCu_2O_{7+\delta}$ superconductor
have been reported by J. Shimoyama {\it et al.} and T. Mochiku {\it et al.}
\cite{21, 22}. This suggests that $Fe_{0.5}Cu_{0.5}Sr_2YCu_2O_{7+\delta}$
superconductor might be obtained. It has been known that R-123 or 1212 phase
can not be prepared by solid-state reaction under ambient pressure in R-Sr-Cu-O
systems. However, in $(Fe_xCu_{1-x})Sr_2YCu_2O_{7+\delta}$ system, single phase
of (Cu,Fe)-1212 phase can be obtained because this phase can be stabilized by
the substitution of $Fe$ for $Cu$.

The $Fe_{0.5}Cu_{0.5}Sr_2YCu_2O_{7+\delta}$ sample prepared under ambient
pressure is not superconducting, but after high pressure synthesis, the sample
becomes superconducting. This suggests that the origin of superconductivity in
$Fe_{0.5}Cu_{0.5}Sr_2YCu_2O_{7+\delta}$ is similar to that in
$Fe_{0.5}Cu_{0.5}BaSrYCu_2O_{7+\delta}$ and
$Fe_{0.5}Cu_{0.5}Ba_2YCu_2O_{7+\delta}$ superconductors of HP-samples. XRD
patterns and $M$-$T$ curve of $Fe_{0.5}Cu_{0.5}Sr_2YCu_2O_{7+\delta}$
superconductor are shown in Fig.7, the $T_c(onset)$ is about 77K, and the
sample is nearly single phase. The lattice parameters for HP-sample are $a$ =
0.3859(4)nm, $c$ = 1.1602(5)nm, while for AM-sample, $a$ = 0.3871(3)nm, $c$ =
1.1669(5)nm.

\subsection{The annealing experiments of
$Fe_{0.5}Cu_{0.5}BaSrYCu_2O_{7+\delta}$ superconductor}

The oxygen content in the $Fe_{0.5}Cu_{0.5}BaSrYCu_2O_{7+\delta}$ samples
prepared by solid-state reaction under ambient pressure and high pressure
synthesis were determined by a volumetric method \cite{43}. The principle of
this method is to dissolve the sample in diluted hydrochloric acid, according
to the liberated oxygen volume, the oxygen content is calculated. The obtained
results are: 7+$\delta$ = 7.20(2) in AM-sample and 7+$\delta$ = 7.35(2) in
HP-sample. This indicates clearly that the oxygen content is increased by high
pressure synthesis.

A superconducting $Fe_{0.5}Cu_{0.5}BaSrYCu_2O_{7+\delta}$ sample with $T_c$
$\sim$ 60K was annealed in air at 210, 290 400, 500, 700, 900 and 950$^o$C for
2 hours step by step.

Fig.8(a) shows the magnetization $vs.$ temperature curves of the annealed
$Fe_{0.5}Cu_{0.5}BaSrYCu_2O_{7+\delta}$ HP-sample, indicating that the
superconducting transition temperature $T_c$ and superconducting volume
fraction decrease rapidly with the increase of annealing temperature, and the
sample became non-superconducting when annealing temperature were higher than
500$^o$C, and even after annealed under as low as 290$^o$C, superconductivity
in this sample was almost destroyed completely. The results of the annealed
$Fe_{0.5}Cu_{0.5}Ba_2YCu_2O_{7+\delta}$ superconductor are similar to these
results, which are shown in Fig.8(b).

Fig.9 shows the local region XRD patterns (46$^o$$\sim$48$^o$ in 2$\theta$) of
the $Fe_{0.5}Cu_{0.5}BaSrYCu_2O_{7+\delta}$ AM-sample and the annealed
HP-sample at different temperatures. After high pressure synthesis, the peaks
of (006) and (200) shift to high angle degree and overlap each other, and with
the increase of annealing temperature, the peaks shift to low angle degree
again. When the annealing temperature is higher than 400$^o$C, the positions of
the peaks are nearly constant within experimental error, indicating that
lattice parameters $a$ and $c$ stop increasing. It means that the extra oxygen
introduced by high pressure synthesis is liberated almost completely at
400$^o$C. As a result, the annealed sample becomes non-superconducting. When
temperature is higher than 500$^o$C, the main role of the annealing is the
improvement of lattice distortion, homogenizing and the increase of crystal
grain size, which makes the peaks narrower and makes (006) and (200) peaks
gradually apart from each other.

\section{Discussions}

A well-known fact in R-123 phase superconductors is that the superconductors
with higher oxygen content have smaller lattice parameters $a$ and $c$. This
phenomenon was also observed in the $YSr_2Cu_{2.7}Fe_{0.3}O_{7+\delta}$
superconductor \cite{20} and in $FeSr_2YCu_2O_{7+\delta}$ superconductor
\cite{22} obtained by high oxygen pressure synthesis. In these superconductors,
the diffraction peaks of (006) and (200) shift  systematically to higher angle
degree with the increase of oxygen content, which is similar to our results.
All our HP-samples have smaller lattice parameters $a$ and $c$ than the
corresponding AM-samples, and the lattice unit cell of HP-samples displays an
obvious shrinkage. Since the relative shrinkage is not more bigger than other
samples annealed under high oxygen pressure, and also for that
$FeSr_2YCu_2O_{7+\delta}$ sample can be made superconducting without high
pressure synthesis, thereby this shrinkage can be attributed to the increase of
oxygen content, but not the pressure effect caused by the high pressure
synthesis procedure. Then it can be deduced that in all
$Fe_{0.5}Cu_{0.5}(Ba_{1-x}Sr_{x})_2YCu_2O_{7+\delta}$ samples the oxygen
content increased after high pressure synthesis, which is validated by the
oxygen content measurement in $Fe_{0.5}Cu_{0.5}BaSrYCu_2O_{7+\delta}$ system
where 7+$\delta$ increased from 7.20(2) in AM-sample to 7.35(2) in HP-sample.
The increased oxygen content provided the required charge which made the
semiconducting-like AM-samples changed to superconducting HP-samples.

The diffusion of $Fe$ atom into $CuO_2$ planes in R-123 systems is believed to
be very unfavorable for superconductivity because of its magnetic moment, and
with the increase of $Fe$ concentration, this diffusion is inevitable in
samples synthesized by common solid-state reaction \cite{PC288}. J. Shimoyama
{\it et al.} \cite{21} used a complex annealing procedure which is believed to
suppress the incorporation of $Fe$ to the $CuO_2$ planes in
$FeSr_2YCu_2O_{7+\delta}$ system, and finally made it superconducting with
$T_c$ $\sim$ 60K. Using neutron powder diffraction studies, T. Mochiku {\it et
al.} \cite{22} investigated the site-mixing between $Fe$ and $Cu$ atoms in
$FeSr_2YCu_2O_{7+\delta}$ samples synthesized through different procedures, and
proved that simply increasing the  oxygen content could not produce
superconductivity and the ordered distribution of $Fe$ and $Cu$ atoms between
$Cu(1)$ and $Cu(2)$ sites is also one of the key points to the origin of
superconductivity. In our heavily $Fe$-doped R-1212 systems, high pressure
synthesis procedure is believed to promote the transfer of $Fe$ atoms from
$Cu(2)$ site to $Cu(1)$ site, and to make almost all the $Fe$ atoms to occupy
the $Cu(1)$ chain site. This redistribution of $Fe$ atoms is a needed
requirement to ensure the occurrence of superconductivity, while the increase
of oxygen content provides the needed amount of charge.

The annealing experiments of $Fe_{0.5}Cu_{0.5}BaSrYCu_2O_{7+\delta}$ and
$Fe_{0.5}Cu_{0.5}Ba_2YCu_2O_{7+\delta}$ superconductors indicate clearly that
the concentration of oxygen content affects superconductivity in HP-samples
directly. Since annealed under the temperature as low as 290$^o$C during a
short time of 2 hours, the occupation site of $Fe$ atoms could not be changed,
but superconductivity in these superconductors was destroyed sharply. The local
region XRD patterns in Fig.9 indicated that the lattice parameters increased
rapidly with the increase of annealing temperature below 400$^o$C, which
suggested the release of oxygen content, and this was the main reason for the
loss of superconductivity.

After the discovery of $FeSr_2YCu_2O_{7+\delta}$ and
$Fe_{0.5}Cu_{0.5}BaSrYCu_2O_{7+\delta}$ superconductors, there is a doubt that
the superconductivity in these superconductors is possibly from
$YBa_2Cu_3O_{7-\delta}$, $YBaSrCu_3O_{7-\delta}$ or $YSr_2Cu_3O_{7-\delta}$
phase formed by phase segregation. It is known that $YSr_2Cu_3O_{7-\delta}$
superconductor with $T_c$ = 30-80K can be obtained also by high pressure
synthesis and its superconductivity disappears after heat treating at
300-500$^o$C. Although the single phase of XRD pattern results and high
superconducting volume fraction of these superconductors have excluded these
disputes, the achievement of $Fe_{0.5}Cu_{0.5}Ba_2YCu_2O_{7+\delta}$
superconductor and the similarity of its superconductivity to those of
$Fe_{0.5}Cu_{0.5}BaSrYCu_2O_{7+\delta}$ and
$Fe_{0.5}Cu_{0.5}Sr_2YCu_2O_{7+\delta}$ superconductors eliminate this doubt.
The annealing experiments also eliminate the possibility that the
superconductivity in $Fe_{0.5}Cu_{0.5}Ba_2YCu_2O_{7+\delta}$ HP-sample is from
$YBa_2Cu_3O_{7-\delta}$ phase due to phase segregation, since it is well known
that the superconductivity in $YBa_2Cu_3O_{7-\delta}$ can not be destroyed by
annealing at the temperature of 200-500$^o$C and the $T_c$ of
$YBa_2Cu_3O_{7-\delta}$ obtained by high pressure synthesis is 92K \cite{46}.

The discovery of these heavily-substituted superconductors with magnetic metal
of $Fe$ atoms provides new platforms to investigate the mechanism of high
temperature superconductivity and the correlation of magnetism and
superconductivity, and it seems that we need to reconsider the influence of
magnetism on superconductivity.

\section{Conclusions}

Three kind of superconductors with high $Fe$ concentration
$Fe_{0.5}Cu_{0.5}Ba_2YCu_2O_{7+\delta}$,
$Fe_{0.5}Cu_{0.5}BaSrYCu_2O_{7+\delta}$ and
$Fe_{0.5}Cu_{0.5}Sr_2YCu_2O_{7+\delta}$ were successfully prepared by
solid-state reaction and high pressure synthesis. The high pressure synthesis
resulted in the increase of oxygen content and redistribution of $Fe$ and $Cu$
atoms between $Cu(1)$ and $Cu(2)$ sites, which are two key factors for the
occurrence of superconductivity. The annealing experimental results indicate
these clearly and eliminate the doubt of superconductivity relative to possible
secondary phase of $YBa_2Cu_3O_{7-\delta}$, $YBaSrCu_3O_{7-\delta}$ or
$YSr_2Cu_3O_{7-\delta}$ superconductors.

\newpage
\centerline{Figure Captions} Fig.1 Typical XRD patterns for AM-samples of
$(Fe_xCu_{1-x})BaSrYCu_2O_{7+\delta}$ system

Fig.2 Lattice parameters $a$, $b$, $c$ and unit cell volume $V$ $vs.$ $x$ in
$(Fe_xCu_{1-x})BaSrYCu_2O_{7+\delta}$ system for AM-samples

Fig.3 Typical $R$-$T$ curves and $T_c$ $vs.$ $x$ in
$(Fe_xCu_{1-x})BaSrYCu_2O_{7+\delta}$ system for AM-samples

Fig.4 $M$-$T$ curves obtained using ZFC mode under the applied field of 10Oe
for HP-samples of $(Fe_xCu_{1-x})BaSrYCu_2O_{7+\delta}$ system; insert shows
$T_c$ $vs.$ $x$

Fig.5 Typical $R$-$T$ and $M$-$T$ curves under the applied field of 10Oe for
$Fe_{0.5}Cu_{0.5}BaSrYCu_2O_{7+\delta}$ HP-sample superconductor

Fig.6 XRD patterns of AM-sample and HP-sample and $R$-$T$ and $M$-$T$ (applied
field of 10Oe) curves of HP-sample of $Fe_{0.5}Cu_{0.5}Ba_2YCu_2O_{7+\delta}$

Fig.7 XRD patterns of AM-sample and HP-sample and $M$-$T$ curves (applied field
of 10Oe) of HP-sample of $Fe_{0.5}Cu_{0.5}Sr_2YCu_2O_{7+\delta}$

Fig.8 (a) $M$-$T$ curves obtained using ZFC mode under applied field of 10Oe
for $Fe_{0.5}Cu_{0.5}BaSrYCu_2O_{7+\delta}$ superconductor annealed at
different temperatures (b) For $Fe_{0.5}Cu_{0.5}Ba_2YCu_2O_{7+\delta}$
superconductor annealed at different temperatures

Fig.9 Local region XRD patterns of $Fe_{0.5}Cu_{0.5}BaSrYCu_2O_{7+\delta}$
AM-sample and HP-sample annealed at different temperatures

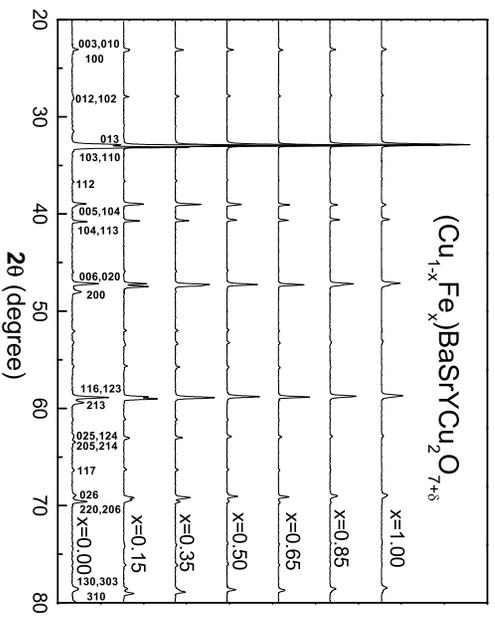
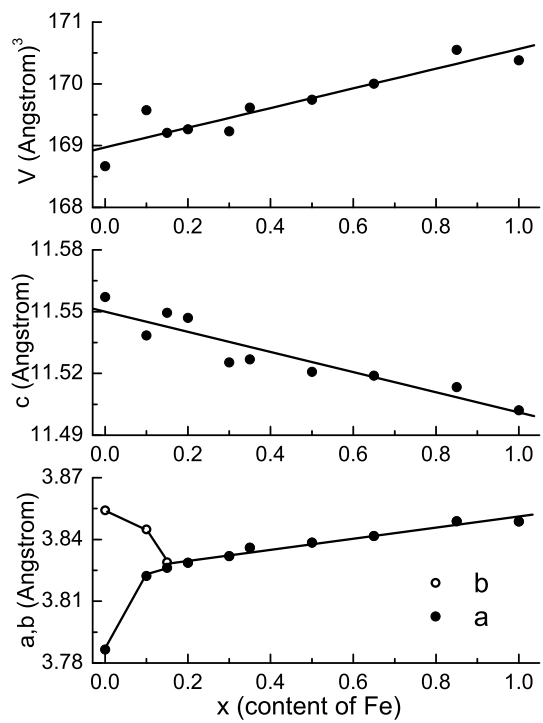
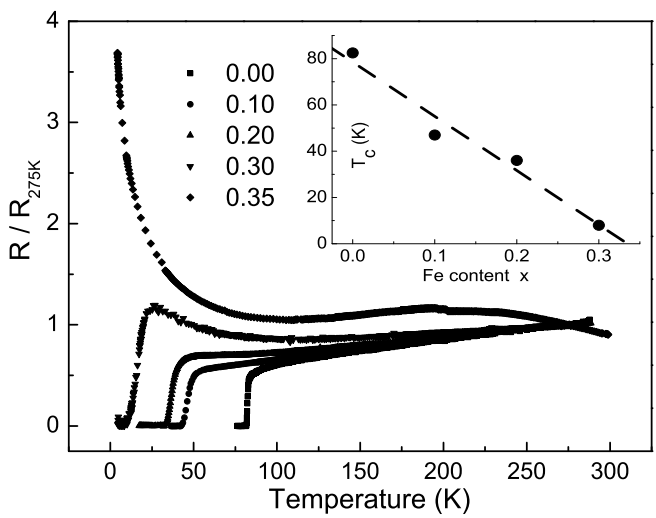
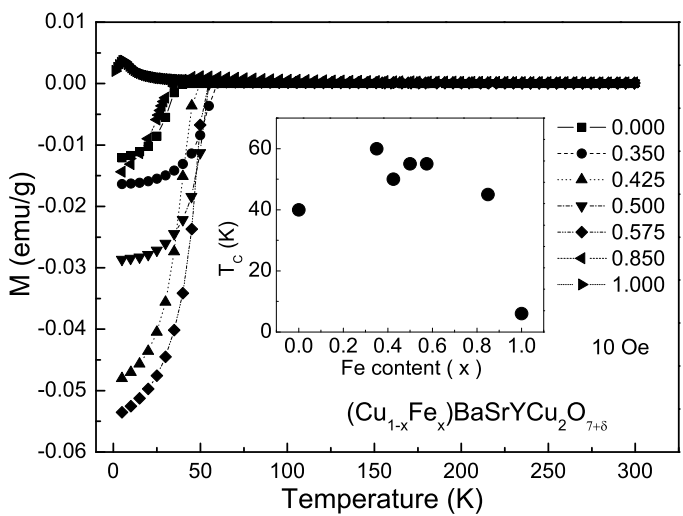
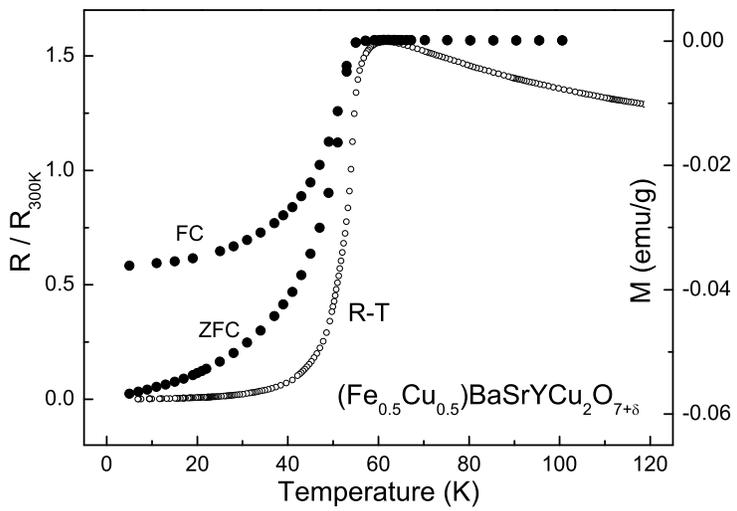
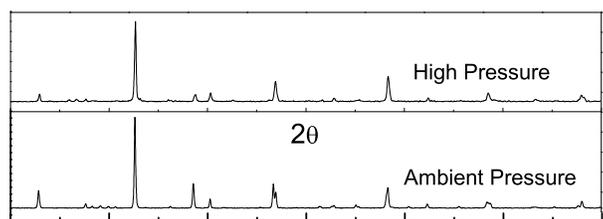
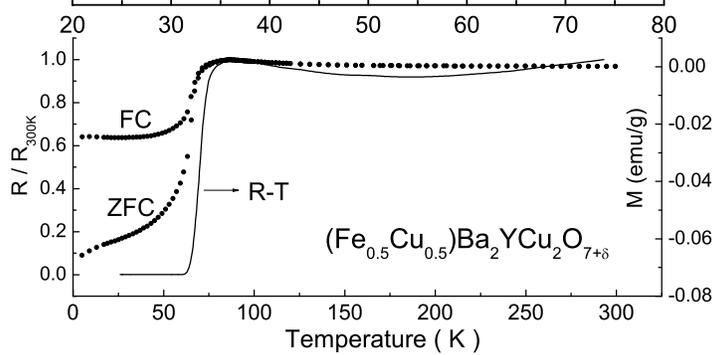
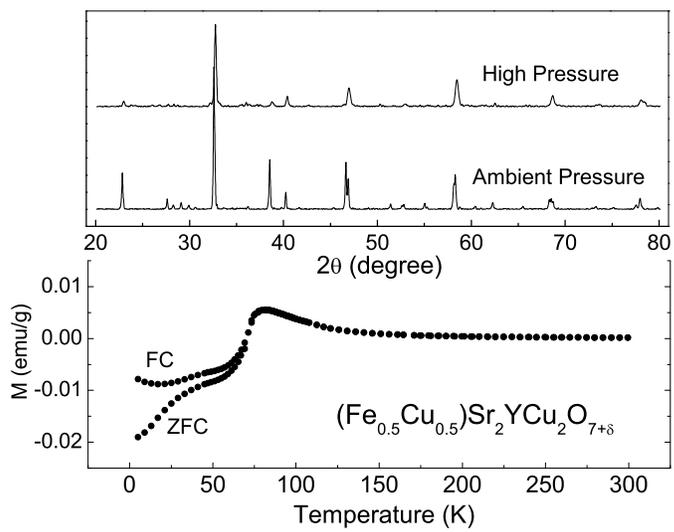
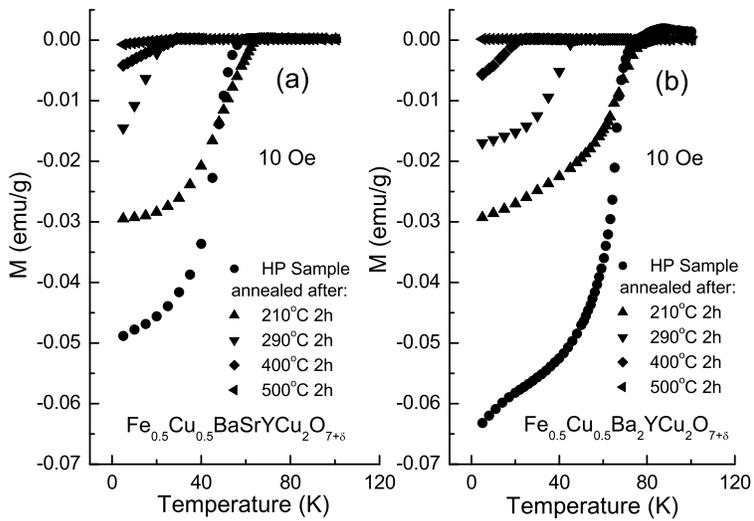
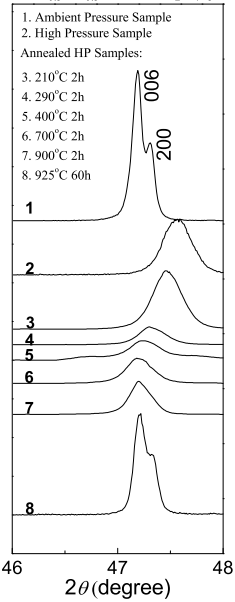
\end{document}